\documentclass{article}
\usepackage[latin1]{inputenc}
\usepackage[english]{babel}
\usepackage{textcomp}
\usepackage[T1]{fontenc}
\usepackage{amsfonts}
\usepackage{amsmath}
\usepackage{amssymb}
\usepackage{mathrsfs}
\usepackage{cite}
\usepackage{graphicx}
\usepackage[small]{caption}[2004/07/16]
\usepackage{braket}
\usepackage{booktabs}

\newtheorem{theorem}{Theorem}

\DeclareMathOperator{\modular}{Sp}

\DeclareMathOperator{\Diag}{Diag}
\DeclareMathOperator{\Imm}{Im}

\DeclareMathOperator{\matr}{M}

\newcommand{\xisei}{\Xi_6[\delta]}
\newcommand{\xiseit}{\Xi_6[\delta](\tau)}
\newcommand{\cc}{\mathbb{C}}

\newcommand{\zz}{\mathbb{Z}}
\newcommand{\hh}{\mathbb{H}}
\newcommand{\pp}{\mathbb{P}}

\newcommand{\be}{\begin{equation}}
\newcommand{\ee}{\end{equation}}
\newcommand{\bes}{\begin{equation*}}
\newcommand{\ees}{\end{equation*}}
\newcommand{\beqs}{\begin{eqnarray*}}
\newcommand{\eeqs}{\end{eqnarray*}}
\newcommand{\de}{\partial}
\newcommand{\der}[2]{{\partial #1\over\partial #2}}
\newcommand{\transp}[1]{{}^t\!#1}
\newcommand{\tthhd}[5]{\theta^{{#5}} \left[
\begin{smallmatrix}
{#1} & {#2}\\
{#3} & {#4}
\end{smallmatrix}
\right]}
\newcommand{\smaq}{\left[ \begin{smallmatrix}}
\newcommand{\smat}{\left( \begin{smallmatrix}}
\newcommand{\smcq}{\end{smallmatrix}\right]}
\newcommand{\smct}{\end{smallmatrix}\right)}
\newcommand{\smag}{\left \{ \begin{smallmatrix}}
\newcommand{\smcg}{\end{smallmatrix}\right \}}

\numberwithin{equation}{section}

\begin{document}
\begin{titlepage}
\vskip 15mm
\begin{center}
\topmargin 30mm
\renewcommand{\thefootnote}{\fnsymbol{footnote}}
{\Large \bf Two loop superstring amplitudes and \boldmath{$S_6$} representations}

\vskip 18mm {\large \bf {
Sergio L.~Cacciatori$^{1,3}$\footnote{sergio.cacciatori@uninsubria.it}
and
Francesco Dalla~Piazza$^{2}$\footnote{francescodp82@yahoo.it}}}\\
\renewcommand{\thefootnote}{\arabic{footnote}}
\setcounter{footnote}{0} \vskip 10mm
{\small $^1$
Dipartimento di Scienze Fisiche e Matematiche, \\
\hspace*{0.15cm} Universit\`a dell'Insubria, \\
\hspace*{0.15cm} Via Valleggio 11, I-22100 Como. \\
\vspace*{0.5cm}

$^2$ Dipartimento di Fisica, \\
Universit\`a degli Studi di Milano,\\
via Celoria 16, I-20133 Milano.\\
\vspace*{0.5cm}

$^3$ INFN, Sezione di Milano, Via Celoria 16, I-20133 Milano.

}
\end{center}
\vspace{0.8cm}
\begin{center}
{\bf Abstract}
\end{center}
{\small {In this paper we describe how representation theory of groups
can be used to shorten the derivation of two loop
partition functions in string theory, giving an intrinsic description of modular forms appearing in the results of
D'Hoker and Phong \cite{hokerIV}. Our method has the advantage of using only algebraic properties of modular functions and it can be extended to any genus $g$.}}

\end{titlepage}


\section{Introduction}
It was conjectured by Belavin and Knizhnik \cite{belavin_alggeom} that ``any multiloop
 amplitude in any conformal invariant string theory may be deduced from
purely algebraic objects on moduli spaces $M_p$ of Riemann surfaces''.
 This was a known fact for zero and for one loop amplitudes.
For bosonic strings, two, three and four loop amplitudes was computed
 (in the same year) in \cite{belavin_twothree,moore,morozov} in terms of modular forms.\\
For superstring theories the story is much longer because of some
 technical difficulties. In particular, the presence of fermionic
 interactions
makes the splitting between chiral and antichiral modes hard. Moreover,
 grassmanian variables arise from worldsheet supersymmetry and
one needs a covariant way to integrate them out. Both problems were
 solved by D'Hoker and Phong, who
in a series of articles \cite{hokerI,hokerII,hokerIII,hokerIV}  showed
 that the computation of $g$-loop string amplitudes in
perturbation theory is strictly connected with the construction of a
 suitable measure on the moduli space of genus $g$ Riemann surfaces.
They claim \cite{hokerasyzI,hokerasyzII} that the genus $g$ vacuum to
 vacuum amplitude must take the form
\be \label{ampiezgenh}
\mathcal{A}=\int_{\mathcal{M}_g}(\det\Imm\tau)^{-5}\sum_{\delta,\bar{\delta}}c_{\delta,\bar{\delta}}d\mu[\delta](\tau)\wedge
\overline{d\mu[\bar{\delta}](\tau)},
\ee
where $\delta$ and $\bar{\delta}$ denote two spin
 structures or theta characteristics, $c_{\delta,\bar{\delta}}$ are
suitable constant phases depending on the details of the model
and $d\mu[\delta](\tau)$ is a holomorphic form of maximal rank
 $(3g-3,0)$ on the moduli space of genus $g$ Riemann surfaces.
The Riemann surface is represented by its period matrix $\tau$, after a choice
 of canonical homology basis.
Since the integrand should be independent from the choice of homology
 basis, it follows that the measure $d\mu[\delta](\tau)$ must transform
covariantly under the modular group $\modular(2g,\zz)$.

In \cite{hokerIV} D'Hoker and Phong have given
an explicit expression for the two loop
 measure in terms of theta constants, i.e. theta functions evaluated at the
 origin, $z=0$. The amplitude \eqref{ampiezgenh} is written
 in terms of modular forms and is manifestly modular
invariant:
\be \label{misura}
d\mu[\delta](\tau)=\frac{\theta^4[\delta](\tau,0)\Xi_6[\delta](\tau,0)}{16\pi^6\Psi_{10}(\tau)}\prod_{I\leq
 J}d\tau_{IJ}.
\ee
Here $\Psi_{10}(\tau)$ is a modular form of weight ten:
\be \label{psi}
\Psi_{10}=\prod_\delta\theta^2[\delta](\tau,0),
\ee
where $\delta$ varies on the whole set of even spin structures
 (consisting of ten elements).
The ten $\xisei$ are defined\footnote{Comparing our conventions with
 the ones of D'Hoker and Phong one should note that our spin matrices are
transposed, according with our conventions on theta functions,
 signatures, etc.} by
\be \label{xi}
\Xi_6[\delta](\tau,0):=\sum_{1\leq
 i<j\leq3}\langle\nu_i|\nu_j\rangle\prod_{k=4,5,6}\theta^4[\nu_i+\nu_j+\nu_k](\tau,0)\ ,
\ee
where each even spin structure is written as a sum of three distinct
 odd spin structures $\delta=\nu_1+\nu_2+\nu_3$ and $\nu_4,\nu_5,\nu_6$
denote the remaining three distinct odd spin structures, see Appendix
 \ref{app:uno}.
The signature of a pair of spin structures, even or odd, is defined by:
\begin{align} \label{segnat}
\langle\kappa|\lambda\rangle:=e^{\pi i(a_\kappa\cdot
 b_\lambda-b_\kappa\cdot a_\lambda)},&&\kappa=
\smaq a_\kappa \\ b_\kappa \smcq, &&\lambda=\smaq a_\lambda \\
 b_\lambda \smcq.
\end{align}
In what follows we will refer to the theta constants as
  $\theta[\delta]:=\theta[\delta](\tau,0)$
and similar for $\xisei$.

Our aim in this letter is to give an intrinsic description of the kind
 of modular forms appearing in two loop amplitudes, and to show how
to give explicit
 expressions of them in terms of theta constants employing group representation techniques . Our method has the advantage of using only algebraic properties of modular functions (in the spirit of
 \cite{belavin_alggeom}) and it can be extended to any genus $g$. In particular it can be used
 to
overcome the difficulties encountered in \cite{hokerasyzI,hokerasyzII}
 for the computation of three loop amplitudes, as will be shown in a forthcoming paper \cite{cvd}.

\section{The Igusa quartic and the forms
 \boldmath{$\xisei$}}\label{sec:xisei}
At genus two, there are ten even spin structures which correspond to
ten theta functions with even characteristics. To study even powers of these functions we define:
\be
\Theta[\varepsilon](\tau)=\theta\smaq \varepsilon \\ 0 \smcq(2\tau,0),
\ee
with $[\varepsilon]=[\varepsilon_1\,\varepsilon_2]$ and we use the formula \cite{fay}:
\be \label{gsvilf}
\theta\smaq \alpha \\ \beta+\gamma \smcq(\tau,z_1+z_2)\theta\smaq \alpha \\ \beta \smcq(\tau,z_1-z_2)=\sum_{\delta\in(\zz/2\zz)^g}(-1)^{\beta\cdot\delta}\theta\smaq \delta \\ \gamma \smcq(2\tau,2z_1)\theta\smaq \alpha+\delta \\ \gamma \smcq(2\tau,2z_2),
\ee
with $z_1=z_2=0$, $\gamma=0$ and $g=2$.
It follows that the fourth powers of the theta functions $\theta[\delta](\tau,z)$, evaluated at the origin, $z=0$, form a five dimensional vector space,
that we call $V_\theta$. We can choose a basis for this space of holomorphic functions on the Siegel space for $g=2$ and, for
 our purpose, a convenient one is:
\begin{align*}
\ P_0 &= \Theta^4\smaq 0 & 0 \smcq+\Theta^4\smaq 0 & 1 \smcq +\Theta^4\smaq 1 & 0 \smcq+\Theta^4\smaq 1 & 1
 \smcq \\
\ P_1 &= 2(\Theta^2\smaq 0 & 0 \smcq\Theta^2\smaq 0 & 1 \smcq+\Theta^2\smaq 1 & 0 \smcq\Theta^2\smaq 1 & 1 \smcq) \\
\ P_2 &= 2(\Theta^2\smaq 0 & 0 \smcq\Theta^2\smaq 1 & 0 \smcq+\Theta^2\smaq 0 & 1 \smcq\Theta^2\smaq 1 & 1 \smcq) \\
\ P_3 &= 2(\Theta^2\smaq 0 & 0 \smcq\Theta^2\smaq 1 & 1 \smcq+\Theta^2\smaq 0 & 1 \smcq\Theta^2\smaq 1 & 0 \smcq) \\
\ P_4 &= 4\Theta\smaq 0 & 0 \smcq\Theta\smaq 0 & 1 \smcq\Theta\smaq 1 & 0 \smcq\Theta\smaq 1 & 1 \smcq,
\end{align*}
The
 expansions of the theta constants on this basis are summarized in Table
 \ref{tab:basP}.
\begin{table}[hbtp]
\begin{center}
\bes
\begin{array}{ccccccc}
\toprule
\delta & \theta^4[\delta] & P_0 & P_1 & P_2 & P_3 & P_4 \\
\midrule
\delta_1 & \tthhd{0}{0}{0}{0}{4} & 1 & 1 & 1 & 1 & 0 \\
\delta_2 & \tthhd{0}{0}{0}{1}{4} & 1 & -1 & 1 & -1 & 0 \\
\delta_3 & \tthhd{0}{0}{1}{0}{4} & 1 & 1 & -1 & -1 & 0\\
\delta_4 & \tthhd{0}{0}{1}{1}{4} & 1 & -1 & -1 & 1 & 0\\
\delta_5 & \tthhd{0}{1}{0}{0}{4} & 0 & 2 & 0 & 0 & 2\\
\delta_6 & \tthhd{0}{1}{1}{0}{4} & 0 & 2 & 0 & 0 & -2\\
\delta_7 & \tthhd{1}{0}{0}{0}{4} & 0 & 0 & 2 & 0 & 2\\
\delta_8 & \tthhd{1}{0}{0}{1}{4} & 0 & 0 & 2 & 0 & -2\\
\delta_9 & \tthhd{1}{1}{0}{0}{4} & 0 & 0 & 0 & 2 & 2\\
\delta_{10} & \tthhd{1}{1}{1}{1}{4} & 0 & 0 & 0 & 2 & -2 \\
\bottomrule
\end{array}
\ees
\end{center}
\caption{Expansion of $\theta^4[\delta]$ on the basis of $P_i$}
 \label{tab:basP}
\end{table}

The period matrix $\tau$, that defines the Riemann surface, at genus
 two belongs to the complex variety
\(\hh_2=\{\tau\in\matr_2(\cc) \mbox{ t.c.: }
 \transp{\tau}=\tau,\;\Imm(\tau)>0\}\). The selected basis defines the map:
\beqs
\varphi_4 : \hh_2 & \stackrel{}{\longrightarrow} & \pp^4 \\
\tau\ & \longmapsto &
 (P_0(\tau):P_1(\tau):P_2(\tau):P_3(\tau):P_4(\tau)).
\eeqs
The closure of the image of \(\varphi_4\) is the ``Igusa quartic'', the
 vanishing locus of
\be \label{igusa}
I_4=P_4^4 + P_4^2P_0^2 - P_4^2P_1^2 - P_4^2P_2^2 - P_4^2P_3^2 +
 P_1^2P_2^2 + P_1^2P_3^2 + P_2^2P_3^2 - 2P_0P_1P_2P_3
\ee
in $\pp^4$.
It is indeed immediate to verify, expressing the $P_i$ in terms of the
 four theta constants $\Theta[\varepsilon]$, that this polynomial is identically zero. We can also write $I_4$ as:
\be \label{igusath}
I_4=\frac{1}{192}\left[\left(\sum_\delta\theta^8[\delta]\right)^2-4\sum_\delta\theta^{16}[\delta]\right].
\ee
We want to find a connection between the forms $\xisei$ appearing in
 the works of D'Hoker and Phong and the Igusa quartic
whose mathematical structure is well known.
For this purpose, we start considering two vector spaces which we call
 $V_\Xi$ and \(V_{\partial_pI}\).
The first one is the space generated by the ten forms \(\xisei\):
\be
V_\Xi=\langle\cdots,\xisei,\cdots\rangle.
\ee
We will see that it is a five dimensional space.
The second vector space we are interested in is the space of the
 derivatives of the Igusa quartic with respect to $P_i$:
\be
V_{\partial_PI}=\langle\cdots,\frac{\partial I_4}{\partial
 P_i},\cdots\rangle_{i=0,\cdots,4},
\ee
which is again a five dimensional space. Both spaces are generated by
 homogeneous polynomials of degree twelve in the theta constants $\Theta[\varepsilon]$ or,
equivalently, of degree three in the $P_i$. We find:
\begin{theorem}
We have \(V_\Xi=V_{\de_{PI}}\), in particular $\dim V_\Xi=5$ and Table \ref{tab:xiseidpi} gives the expansion of each $\Xi_6[\delta]$ as linear combination of the
 derivative of Igusa quartic with respect to $P_i$.
\end{theorem}
\begin{table}[htbp]
\begin{center}
\bes
\begin{array}{cccccc}
\toprule
\delta & \partial_{P_0} I_4 & \partial_{P_1} I_4 & \partial_{P_2} I_4 &
 \partial_{P_3} I_4 & \partial_{P_4} I_4 \\
\midrule
\Xi_6[\delta_1] & 6 & 2 & 2 & 2 & 0 \\
\Xi_6[\delta_2] & 6 & -2 & 2 & -2 & 0 \\
\Xi_6[\delta_3] & 6 & 2 & -2 & -2 & 0 \\
\Xi_6[\delta_4] & 6 & -2 & -2 & 2 & 0 \\
\Xi_6[\delta_5] & 0 & 4 & 0 & 0 & 2 \\
\Xi_6[\delta_6] & 0 & 4 & 0 & 0 & -2 \\
\Xi_6[\delta_7] & 0 & 0 & 4 & 0 & 2 \\
\Xi_6[\delta_8] & 0 & 0 & 4 & 0 & -2 \\
\Xi_6[\delta_9] & 0 & 0 & 0 & 4 & 2 \\
\Xi_6[\delta_{10}] & 0 & 0 & 0 & 4 & -2 \\
\bottomrule
\end{array}
\ees
\end{center}
\caption{Expansion of the functions $\xiseit$ on the $\frac{\partial
 I_4}{\partial P_i}$. We intend $\partial_{P_0} I_4\equiv\frac{\partial
 I_4}{\partial P_i}$.}
\label{tab:xiseidpi}
\end{table}

Another interesting vector space is the one generated by the
 derivatives of the Igusa quartic with respect to the ten theta
constants $\theta[\delta]$ at the fourth power:
\be
V_{\partial_\theta I}:=\langle\cdots,\frac{\partial I_4}{\partial
 \theta^4[\delta]},\cdots\rangle.
\ee
In computing these derivatives the theta constants \(\theta^4[\delta]\)
 must be considered as independent functions and we use \eqref{igusath}.
$V_{\partial_\theta I}$ has dimension ten, so these polynomials are all
 independent. Next define the ten functions:
\be \label{fstr}
f_\delta:=2\xisei-\frac{\partial I_4}{\partial \theta^4[\delta]},
\ee
generating the vector space $V_f=\langle\cdots,f_\delta,\cdots\rangle$
 of dimension five. Then:
\begin{align} \label{decV}
& \sum_\delta\der{I_4}{\theta^4[\delta]}f_\delta=0 &\mbox{and}&& V_{\partial_\theta I}=V_f\oplus V_\Xi.
\end{align}

This connection of the Igusa quartic with the forms \(\xisei\)
 suggests studying the whole space of the polynomials of degree three in the
 $P_i$:
$S^3V_\theta=\langle\cdots,P_iP_jP_k,\cdots\rangle_{0\leq i\leq j\leq
 k\leq 4}$, the triple symmetric tensor product of the space $V_\theta$.
We want to decompose this $35$ dimensional space in a ``natural'' way
 and understand which parts of such a decomposition are involved in the measure \eqref{misura}.

\section{Decomposition of \boldmath{$S^3V_\theta$}}\label{sec:s3vtheta}
To decompose the whole space $S^3V_\theta$ in a ``natural'' way as a
 direct sum of vector spaces, $S^3V_\theta=\bigoplus_iV_i$,
we employ the theory of representations of finite groups.
The point is that string amplitudes must be invariant under the action
 of the modular group \(\modular(2g,\zz)\).
In particular for genus two surfaces the modular group is
 \(\modular(4,\zz)\equiv\Gamma_2\).
This group can be surjectively mapped into
the symmetric group \(S_6\) with kernel
 \(\Gamma_2(2)=\{M\in\Gamma_2,\;\; M\equiv Id \pmod{2}\}\), so that
\(S_6\simeq \Gamma_2/\Gamma_2(2)\).
The action of \(S_6\) on the theta constants $\theta^4[\delta]$
 together with the representation theory of finite groups provide the
tools to understand how the space \(S^3V_\theta\) decomposes in terms of
 invariant subspaces under the action of the modular group and which
combinations of theta constants generate each subspace.

To study the action of the symmetric group $S_6$ on $V_\theta$ we have
 to relate the generators of the modular group, see Appendix
 \ref{app:due},
to the elements of $S_6$. We report this relation in Table
 \ref{tab:modsym}.
\begin{table}[htbp]
\begin{center}
\bes
\begin{array}{cccccc}
\toprule
M_1 & M_2 & M_3 & S & \Sigma & T \\
\midrule
(1\,3) & (2\,4) & (1\,3)(2\,4)(5\,6) & (3\,5)(4\,6) &
 (1\,2)(3\,4)(5\,6) & (1\,3)(2\,6)(4\,5) \\
\bottomrule
\end{array}
\ees
\end{center}
\caption{Relationship between the generators of the modular group and
 $S_6$.} \label{tab:modsym}
\end{table}
Each generator induces a permutation of the six odd characteristics $\nu_1,\cdots,\nu_6$ and thus defines an element of $S_6$.
Writing the even characteristics as sum of three odd characteristics, as explained
 in Appendix \ref{app:uno}, we find how the even theta constants $\theta^4[\delta]$
transform under the action of $\modular(4,\zz)$.

We want to identify the representation of $S_6$ on $V_\theta$. This can be obtained fixing
a basis for $V_\theta$, for example $\theta^4[\delta_1]$,
 $\theta^4[\delta_2]$, $\theta^4[\delta_3]$, $\theta^4[\delta_4]$,
 $\theta^4[\delta_5]$,
to compute the representation matrices of $M_i$, $S$, $\Sigma$ and $T$
 and thus of the generators of $S_6$.
The symmetric group $S_6$ has eleven conjugacy classes and thus has eleven irreducible representations, as shown in Table
\ref{tab:caratteri}.
\begin{table}[htbp]
\begin{center}
\bes
\begin{array}{ccrrrrrrrrrrr} \toprule
S_6 & \mbox{Partition} & C_1 & C_2 & C_3 & C_{2,2} & C_4 & C_{3,2} & C_{5} & C_{2,2,2} &
 C_{3,3} & C_{4,2} & C_6 \\
\midrule
\mbox{id}_1 & [6] & 1 & 1 & 1 & 1 & 1 & 1 & 1 & 1 & 1 & 1 & 1 \\
\mbox{alt}_1 & [1^6] & 1 & -1 & 1 & 1 & -1 & -1 & 1 & -1 & 1 & 1 & -1 \\
\mbox{st}_5 & [2^3] & 5 & -1 & -1 & 1 & 1 & -1 & 0 & 3 & 2 & -1 & 0 \\
\mbox{sta}_5 & [3^2] & 5 & 1 & -1 & 1 & -1 & 1 & 0 & -3 & 2 & -1 & 0 \\
\mbox{rep}_5 & [5\,1] &  5 & 3 & 2 & 1 & 1 & 0 & 0 & -1 & -1 & -1 & -1 \\
\mbox{repa}_5 & [2\,1^4] & 5 & -3 & 2 & 1 & -1 & 0 & 0 & 1 & -1 & -1 & 1 \\
\mbox{n}_{9} & [4\,2] & 9 & 3 & 0 & 1 & -1 & 0 & -1 & 3 & 0 & 1 & 0 \\
\mbox{na}_{9} & [2^2\,1^2] & 9 & -3 & 0 & 1 & 1 & 0 & -1 & -3 & 0 & 1 & 0 \\
\mbox{sw}_{10} & [3\,1^3] & 10 & -2 & 1 & -2 & 0 & 1 & 0 & 2 & 1 & 0 & -1 \\
\mbox{swa}_{10} & [4\,1^2] & 10 & 2 & 1 & -2 & 0 & -1 & 0 & -2 & 1 & 0 & 1 \\
\mbox{s}_{16} & [3\,2\,1] & 16 & 0 & -2 & 0 & 0 & 0 & 1 & 0 & -2 & 0 & 0 \\
\bottomrule
\end{array}
\ees
\end{center}
\caption{Characters of the conjugacy classes of the eleven irreducible
 representations of $S_6$.} \label{tab:caratteri}
\end{table}
For example, the conjugacy class $C_{3,2}$ consists of the product of a
 2-cycle and a 3-cycle and the character of the first 10 dimensional representation, $\mathrm{sw}_{10}$, for this class is 1.
The space $V_\theta$ is five dimensional, therefore it must be one of
 the four representations of this dimension.
Looking at the character of the matrix representing $M_1$ allows us to
 identify $V_\theta$ with $\mathrm{st}_5$.

An alternative way to reach the same result is provideed by the Thomae
 formula \cite{fay,mumfordII}:
\be \label{thomae}
\theta^4[\delta]=c\,\epsilon_{S,T}\prod_{i,j\in S\;i<j}(u_i-u_j)
\prod_{k,l\in T\;k<l}(u_k-u_l),
\ee
where \(u_i\) are the six branch points of the Riemann surface of genus
 two, $S$ and $T$ contain the indices of the odd characteristics in the
 two
triads which yield the same even characteristic\footnote{For example for
 \(\delta_4\), \(S=\{1,4,5\}\) and \(T=\{2,3,6\}\).}, as explained in
\cite{hokerIV} or \cite{rauch2},  $\epsilon_{S,T}$ is a sign depending
 on the triads, as indicated in Table \ref{tab:sign}, and $c$ is a
 constant
independent from the characteristic.
\begin{table}[htbp]
\begin{center}
\bes
\begin{array}{cccccccccccccccccccc}
\toprule
146 & 126 & 125 & 145 & 124 & 156 & 123 & 134 & 136 & 135 \\
235 & 345 & 346 & 236 & 356 & 234 & 456 & 256 & 245 & 246 \\
\midrule
\delta_1 & \delta_2 & \delta_3 & \delta_4 & \delta_5 & \delta_6 &
 \delta_7 &\delta_8 & \delta_9 & \delta_{10} \\
-1 & 1 & 1 & -1 & 1 & -1 & 1 & -1 & -1 &-1 \\
\bottomrule
\end{array}
\ees
\end{center}
\caption{Relative signs between the theta constants $\theta^4[\delta]$
 for the Thomea formula.}
\label{tab:sign}
\end{table}
The Thomae formula shows that $S_6$ acts on the theta constants by
 permuting the branch points. Evaluating in this way the effect of
 permutations,
and comparing the characters we find again that the representation
 $V_\theta$ must be identified with $\mathrm{st}_5$.

Thus the representation on the space $S^3V_\theta$ is the
 $\mathrm{S}^3(\mathrm{st}_5)$ that decomposes as follows:
\be \label{decompst}
\mathrm{S}^3(\mathrm{st}_5)=\mathrm{id}_1+\mathrm{n}_9+\mathrm{repa}_5+2\mathrm{st}_5+\mathrm{sw}_{10}.
\ee
The presence of $\mathrm{id}_1$, the trivial representation of $S_6$, implies the
 existence of an invariant polynomial. Its expression in terms of the basis
 $P_i$,
up to a scalar, is:
\be \label{cubinv}
\Psi_6=P_0^3-9P_0(P_1^2+P_2^2+P_3^2-4P_4^2)+54P_1P_2P_3,
\ee
and essentially it is the modular form of weight six appearing in
\cite{hokerIV}.

We will now identify some subspaces of $S^3V_\theta$ in the
 decomposition  \eqref{decompst}.
All these subspaces must be invariant over the action of the modular
 group otherwise a modular transformation of $\theta^4[\delta]$ would
send an element of a subspace in another one. We summarize the
 results in Table \ref{tab:subsp}.
\begin{table}[!h]
\begin{center}
\resizebox*{0.95\textwidth}{!}{
\begin{tabular}{lcc}
\toprule
Space & Dimension & Representation \\
\midrule
$\boldsymbol{ \langle P_0^3+\cdots+54P_1P_2P_3\rangle\equiv V_I}$ &
 {\bf 1} & $\boldsymbol{ \mathrm{id}_1}$ \\
$\boldsymbol{ \langle\partial_{P_i}I_4
 \rangle\equiv\langle\xisei\rangle\equiv V_\Xi}$ & {\bf 5} & $\boldsymbol{ \mathrm{st}_5 }$\\
$\boldsymbol{ \langle 2\xisei-\frac{\partial I_4}{\partial
 \theta^4[\delta]}\rangle\equiv V_f }$ & {\bf 5} & $\boldsymbol{ \mathrm{repa}_5}$
 \\
$\boldsymbol{ \langle \theta^4[\delta_i]\sum_{\delta'}\theta^8[\delta']
 \rangle\equiv V_S}$ & {\bf 5} & $\boldsymbol{\mathrm{st}_5}$\\
$\langle\frac{\partial I_4}{\partial \theta^4[\delta_i]}\rangle$ & 10 &
 $\mathrm{st}_5 \oplus \mathrm{repa}_5$\\
$\langle \theta^{12}[\delta_i] \rangle$ & 10 & $\mathrm{st}_5 \oplus
 \mathrm{repa}_5$\\
$\langle\theta^{12}[\delta_i],\frac{\partial I_4}{\partial
 \theta^4[\delta_j]}\rangle$ & 15 & 2$\mathrm{st}_5 \oplus
 \mathrm{repa}_5$\\
$\langle\theta^{12}[\delta_i],\xisei\rangle$ & 15 & 2$\mathrm{st}_5
 \oplus \mathrm{repa}_5$\\
$\langle
 \theta^{12}[\delta_i],\theta^4[\delta_j]\sum_{\delta'}\theta^8[\delta'] \rangle$ & 15 & 2$\mathrm{st}_5 \oplus \mathrm{repa}_5$\\
$\langle
 \theta^{12}[\delta_i],\theta^4[\delta_j]\sum_{\delta'}\theta^8[\delta'],\partial_{\delta_k}I_4 \rangle$ & 15 & 2$\mathrm{st}_5 \oplus
 \mathrm{repa}_5$\\
$\boldsymbol{ \langle
 \theta^4[\delta_i]\theta^4[\delta_j]\theta^4[\delta_k]\rangle_{\delta_i+\delta_j+\delta_k\mbox{\scriptsize\ odd}}}$
& {\bf 20} & $\boldsymbol{ \mathrm{st}_5 \oplus \mathrm{repa}_5\oplus
 \mathrm{sw}_{10}}$\\
$\boldsymbol{ \langle\theta^4[\delta_i]\theta^8[\delta_j]\rangle}$ &
 {\bf 34} & $\boldsymbol{ 2\mathrm{st}_5 \oplus \mathrm{repa}_5 \oplus
 \mathrm{n}_9\oplus \mathrm{sw}_{10}}$\\
$\langle
 \theta^4[\delta_i]\theta^4[\delta_j]\theta^4[\delta_k]\rangle_{\delta_i,\delta_j,\delta_k\mbox{\scriptsize\ even}}$ & 35 &
 $S^3V_\theta$\\
$\langle
 \theta^4[\delta_i]\theta^4[\delta_j]\theta^4[\delta_k]\rangle_{\delta_i+\delta_j+\delta_k\mbox{\scriptsize\ even}}$ & 35 &
 $S^3V_\theta$\\
\bottomrule
\end{tabular}
}
\caption{Decomposition of the given subspaces}
\label{tab:subsp}
\end{center}
\end{table}
\newline
The final decomposition of the whole space $S^3V_\theta$ is then:
\be \label{decs3v}
S^3V_\theta=V_I \oplus V_\Xi \oplus V_f \oplus V_S \oplus V_9 \oplus
 V_{10},
\ee
where $V_I$ is the subspace generated by the invariant polynomial
 $\Psi_6$ \eqref{cubinv}, $V_\Xi$ is generated by the forms $\xisei$,
$V_f$ is generated by the functions defined in \eqref{fstr} and $V_9$
 and $V_{10}$ are parts of the subspaces of dimension 20 or 34 given
in Table \ref{tab:subsp}.
\newline
Note that $\Psi_6$ can't be written as a linear combination of the products
$\theta^4[\delta_i]\theta^4[\delta_j]\theta^4[\delta_k]$ for
$\delta_i+\delta_j+\delta_k$ an odd characteristic, in contradiction to the claim in
 \cite{hoker0}, because the subspace $V_I$ is not contained in
\(\langle
 \theta^4[\delta_i]\theta^4[\delta_j]\theta^4[\delta_k]\rangle_{\delta_i+\delta_j+\delta_k\mbox{\scriptsize \ odd}}\). Instead $\Psi_6$ can be written as a linear combination of the products $\theta^4[\delta_i]\theta^4[\delta_j]\theta^4[\delta_k]$ for
$\delta_i+\delta_j+\delta_k$ an even characteristic, as correctly said in \cite{hokerIV}. Indeed these products of theta constants span the whole $S^3V_\theta$.

\section{Conclusions}

In this letter we clarified the algebraic properties of the modular
 structures underlying two loop superstring amplitudes.
In the papers of D'Hoker and Phong it was shown that the
 crucial ingredients are the modular forms $\xisei$ appearing
in \eqref{misura}.
In section \ref{sec:xisei} we have connected the forms $\xisei$ to the
 mathematically well known Igusa quartic. This clarifies the
origin of such forms which result to live in a given five dimensional
 subspace of the vector space of cubic polynomials in the fourth powers of the 10 even theta constants. We studied the whole space in Section
 \ref{sec:s3vtheta} where we decomposed it in irreducible representations (irreps)
of the group $S_6$, a quotient of the modular group. In this way we
 identified the irrep corresponding to the space generated by the
forms $\xisei$. Our analysis can be extended to any genus $g$ and gives a direct and quick strategy for searching modular forms with certain properties. 
However, there are some difficulties in carrying on such a generalization.
Possibly equation \eqref{ampiezgenh} is no more true for genus $g>2$ for the following reasons \cite{witten_comm}: D'Hoker and Phong obtained \eqref{ampiezgenh} from a chiral splitting which works using the fact that, for a $g=2$ super Riemann surface with an even spin structure, there are two even holomorphic differentials and no odd ones. The second point necessary for the splitting is that by taking the periods of the two holomorphic differentials, one associates to the original super Riemann surface $M$ an abelian variety $J$, so that one maps the given super Riemann surface $M$ to the ordinary Riemann surface $M'$ that has $M$ for its Jacobian. For a $g>2$ super Riemann surface with an even spin structure there are ``generically'' $g$ even holomorphic differentials and no odd ones, but it is possible to have odd ones for special complex structures on $M$.
So, in an arbitrary genus $g$ where we can have also odd holomorphic differentials, this procedure can not be carried on. Also, if there are no odd holomorphic differentials, taking the periods of the even holomorphic differentials will give us an abelian variety, but it won't necessarily be the Jacobian of an ordinary Riemann surface. Its period can differ from those of an arbitrary Riemann surface by terms that are bilinear in fermionic moduli. Thus equation \eqref{ampiezgenh} requires an improvement for $g>2$. 

Such issue and similar, together with the application of our analysis to the construction of genus three amplitudes \cite{cvd}
and to open and type $O$ string amplitudes will be the
 goals of future papers.

\vskip 1.5 cm
\subsection*{Acknowledgments}
We are grateful to Bert Van Geemen for the idea which underlies this work and for several stimulating discussions.
We are indebted with Edward Witten for explaining us possible difficulties, which we reported in the conclusions, to extend \eqref{ampiezgenh} for higher genus.
We would also like to thank Silvia Manini for suggestions.\\
This work was partially supported by INFN.

\appendix
\section{Spin structure} \label{app:uno}
At genus two there are sixteen independent characteristics, six odd and
 ten even. The odd characteristics are:
\begin{eqnarray*}
\nu_1=
\begin{bmatrix}
0 & 1 \\
0 & 1 \\
\end{bmatrix},
\ \nu_2=
\begin{bmatrix}
1 & 0 \\
1 & 0 \\
\end{bmatrix},
\ \nu_3=
\begin{bmatrix}
0 & 1 \\
1 & 1 \\
\end{bmatrix},\
\nu_4=
\begin{bmatrix}
1 & 0 \\
1 & 1 \\
\end{bmatrix},
\ \nu_5=
\begin{bmatrix}
1 & 1 \\
0 & 1 \\
\end{bmatrix},
\ \nu_6=
\begin{bmatrix}
1 & 1 \\
1 & 0 \\
\end{bmatrix}
.
\end{eqnarray*}
The even characteristics are:
\begin{eqnarray*}
\delta_1=
\begin{bmatrix}
0 & 0 \\
0 & 0 \\
\end{bmatrix},\
\delta_2=
\begin{bmatrix}
0 & 0 \\
0 & 1 \\
\end{bmatrix},\
\delta_3=
\begin{bmatrix}
0 & 0 \\
1 & 0 \\
\end{bmatrix},\
\delta_4=
\begin{bmatrix}
0 & 0 \\
1 & 1 \\
\end{bmatrix},\
\delta_5=
\begin{bmatrix}
0 & 1 \\
0 & 0 \\
\end{bmatrix},\\
\delta_6=
\begin{bmatrix}
0 & 1 \\
1 & 0 \\
\end{bmatrix},\
\delta_7=
\begin{bmatrix}
1 & 0 \\
0 & 0 \\
\end{bmatrix},\
\delta_8=
\begin{bmatrix}
1 & 0 \\
0 & 1 \\
\end{bmatrix},\
\delta_9=
\begin{bmatrix}
1 & 1 \\
0 & 0 \\
\end{bmatrix},\
\delta_{10}=
\begin{bmatrix}
1 & 1 \\
1 & 1 \\
\end{bmatrix}.
\end{eqnarray*}
Each even characteristic can be obtained in two distinct way as a sum
 of three odd characteristics \cite{hokerIV,rauch2} as shown in Table
\ref{tab:azgrmod}.
\begin{table}[htbp]
\begin{center}
\bes
\begin{array}{ccccccccccc}
\toprule
\mbox{Triad} &
\begin{array}{c} 146 \\ 235 \end{array} &
\begin{array}{c} 126 \\ 345 \end{array} &
\begin{array}{c} 125 \\ 346 \end{array} &
\begin{array}{c} 145 \\ 236 \end{array} &
\begin{array}{c} 124 \\ 356 \end{array} &
\begin{array}{c} 156 \\ 234 \end{array} &
\begin{array}{c} 123 \\ 456 \end{array} &
\begin{array}{c} 134 \\ 256 \end{array} &
\begin{array}{c} 136 \\ 245 \end{array} &
\begin{array}{c} 135 \\ 246 \end{array} \\
\midrule
\lbrack \delta \rbrack &
\smaq
0 & 0 \\
0 & 0 \\
\smcq &
\smaq
0 & 0 \\
0 & 1 \\
\smcq &
\smaq
0 & 0 \\
1 & 0 \\
\smcq &
\smaq
0 & 0 \\
1 & 1 \\
\smcq &
\smaq
0 & 1 \\
0 & 0 \\
\smcq &
\smaq
0 & 1 \\
1 & 0 \\
\smcq &
\smaq
1 & 0 \\
0 & 0 \\
\smcq &
\smaq
1 & 0 \\
0 & 1 \\
\smcq &
\smaq
1 & 1 \\
0 & 0 \\
\smcq &
\smaq
1 & 1 \\
1 & 1 \\
\smcq \\[.4em]
\delta & \delta_1  & \delta_2  & \delta_3 & \delta_4 & \delta_5 &
 \delta_6 & \delta_7 & \delta_8 & \delta_9 & \delta_{10}\\
\bottomrule
\end{array}
\ees
\end{center}
\caption{Combinations of odd characteristics that form the same even
 characteristic.}
\label{tab:azgrmod}
\end{table}
\\
In the first line are listed the indices of the two sets of three odd
 characteristics that summed give the same even characteristic.

\section{The modular group \boldmath{$\modular(4,\zz)$}}
 \label{app:due}

The modular group $\modular(4,\zz)$ is defined by the matrices $M=\smat
 A & B \\ C & D \smct$ satisfying:
\bes
\left(\begin{array}{cc}
A & B \\
C & D \\
\end{array}\right)
\left(\begin{array}{cc}
0 & I \\
-I & 0 \\
\end{array}\right)^{\ t}
\!\!\! \left(\begin{array}{cc}
A & B \\
C & D \\
\end{array}\right)
=
\left(\begin{array}{cc}
0 & I \\
-I & 0 \\
\end{array}\right),
\ees
where $A$, $B$, $C$, $D\in\matr_2(\zz)$. The group is generated by:
\begin{align*}
M_i &=
\begin{pmatrix}
I & B_i \\
0 & I \\
\end{pmatrix},
&B_1 &=
\begin{pmatrix}
1 & 0 \\
0 & 0 \\
\end{pmatrix},
&B_2 &=
\begin{pmatrix}
0 & 0 \\
0 & 1 \\
\end{pmatrix},
&B_3 &=
\begin{pmatrix}
0 & 1 \\
1 & 0 \\
\end{pmatrix};
&S &=
\begin{pmatrix}
0 & I \\
-I & 0 \\
\end{pmatrix}; \\
\Sigma &=
\begin{pmatrix}
\sigma & 0 \\
0 & -\sigma \\
\end{pmatrix},
&\sigma &=
\begin{pmatrix}
0 & 1 \\
-1 & 0 \\
\end{pmatrix};
&T &=
\begin{pmatrix}
\tau_+ & 0 \\
0 & \tau_{-} \\
\end{pmatrix},
&\tau_+ &=
\begin{pmatrix}
1 & 1 \\
0 & 1 \\
\end{pmatrix},
&\tau_{-} &=
\begin{pmatrix}
1 & 0 \\
-1 & 1 \\
\end{pmatrix}.
\end{align*}
The action of the modular group on a characteristic $\kappa$ (even or
 odd), at genus $g=2$, is given by:
\be
\begin{pmatrix}
\transp{\tilde{a}} \\
\transp{\tilde{b}} \\
\end{pmatrix}
=
\begin{pmatrix}
D & -C \\
-B & A \\
\end{pmatrix}
\begin{pmatrix}
\transp{a} \\ \transp{b} \\
\end{pmatrix} +\frac{1}{2}\Diag
\begin{pmatrix}
C\cdot\transp{D} \\
A\cdot\transp{B} \\
\end{pmatrix},
\ee
where $a$ and $b$ are the rows of the characteristic $\kappa=\smat a \\
 b \smct$. $\Diag(M)$ for a $n\times n$ matrix $M$ is an $1\times n$
 column
vector whose entries are the diagonal entries of $M$.
The action of a modular transformation on a period matrix is
\be \label{modper}
\tilde{\tau}=(A\tau+B)(C\tau+D)^{-1},
\ee
and on the theta functions:
\be \label{modtheta}
\theta[\tilde{\kappa}](\tilde{\tau},\transp{(C\tau+D)^{-1}}z)=\epsilon(\kappa,M)\det(C\tau+D)^{\frac{1}{2}}e^{\pi
 i\transp{z}(C\tau+D)^{-1}Cz}\theta[\kappa](\tau,z).
\ee
The phase factor $\epsilon(\kappa,M)$, satisfying
 $\epsilon^8(\kappa,M)=1$, depends both on the characteristic $\kappa$ and on the matrix $M$
 generating
the transformation. For the even characteristics $\delta=\smat a \\ b
 \smct$ the fourth powers of $\epsilon$ are given by:
\beqs
\ &&\epsilon^4(\delta,M_i)=e^{\pi i\transp{a}B_ia}\;\;\;\;\;\;\;\;i=1,2
 \\
\
 &&\epsilon^4(\delta,M_3)=\epsilon^4(\delta,S)=\epsilon^4(\delta,\Sigma)=\epsilon^4(\delta,T)=1.
\eeqs
The action of the six generators on the theta characteristics and on
 the triads are reported in Table \ref{tab:generat}.
\begin{table}[!h]
\begin{center}
\bes
\begin{array}{ccccccccccc}
\toprule
\mbox{Triad} & [\delta] & \delta & M_1 & M_2 & M_3 & S & \Sigma & T &
 \epsilon^4(\delta,M_1) & \epsilon^4(\delta,M_2) \\
\midrule
146\;235 & \smaq
0 & 0 \\
0 & 0 \\
\smcq
 & \delta_1 & \delta_3 & \delta_2 & \delta_1 & \delta_1 & \delta_1 &
 \delta_1 & + & + \\
126\;345 & \smaq
0 & 0 \\
0 & 1 \\
\smcq
 & \delta_2 & \delta_4 & \delta_1 & \delta_2 & \delta_5 & \delta_3 &
 \delta_4 & + & + \\
125\;346 & \smaq
0 & 0 \\
1 & 0 \\
\smcq
 & \delta_3 & \delta_1 & \delta_4 & \delta_3 & \delta_7 & \delta_2 &
 \delta_3 & + & + \\
145\;236 & \smaq
0 & 0 \\
1 & 1 \\
\smcq
 & \delta_4 & \delta_2 & \delta_3 & \delta_4 & \delta_9 & \delta_4 &
 \delta_2 & + & + \\
124\;356 & \smaq
0 & 1 \\
0 & 0 \\
\smcq
 & \delta_5 & \delta_6 & \delta_5 & \delta_6 & \delta_2 & \delta_7 &
 \delta_5 & + & - \\
156\;234 & \smaq
0 & 1 \\
1 & 0 \\
\smcq
 & \delta_6 & \delta_5 & \delta_6 & \delta_5 & \delta_8 & \delta_8 &
 \delta_9 & + & - \\
123\;456 & \smaq
1 & 0 \\
0 & 0 \\
\smcq & \delta_7 & \delta_7 & \delta_8 & \delta_8 & \delta_3 & \delta_5
 & \delta_9 & - & + \\
134\;256 & \smaq
1 & 0 \\
0 & 1 \\
\smcq
 & \delta_8 & \delta_8 & \delta_7 & \delta_7 & \delta_6 & \delta_6 &
 \delta_{10} & - & + \\
136\;245 & \smaq
1 & 1 \\
0 & 0 \\
\smcq
 & \delta_9 & \delta_9 & \delta_9 & \delta_{10} & \delta_4 & \delta_9 &
 \delta_7 & - & - \\
135\;246 & \smaq
1 & 1 \\
1 & 1 \\
\smcq
 & \delta_{10} & \delta_{10} & \delta_{10} & \delta_9 & \delta_{10} &
 \delta_{10} & \delta_8 & - & - \\
\bottomrule
\end{array}
\ees
\end{center}
\caption{Transformation of the even characteristics under the action of
 the modular group.}
\label{tab:generat}
\end{table}
\\ In Table \ref{tab:actodd} we report the action of the generators of
 the modular group on the odd characteristics.
\begin{table}[!h]
\begin{center}
\bes
\begin{array}{cccccccc}
\toprule
\left[\nu\right] & \nu & M_1 & M_2 & M_3 & S & \Sigma & T \\
\midrule
\smaq
0 & 1 \\
0 & 1 \\
\smcq
 & \nu_1 & \nu_3 & \nu_1 & \nu_3 & \nu_1 & \nu_2 & \nu_3 \\
\smaq
1 & 0 \\
1 & 0 \\
\smcq
 & \nu_2 & \nu_2 & \nu_4 & \nu_4 & \nu_2 & \nu_1 & \nu_6 \\
\smaq
0 & 1 \\
1 & 1 \\
\smcq
 & \nu_3 & \nu_1 & \nu_3 & \nu_1 & \nu_5 & \nu_4 & \nu_1 \\
\smaq
1 & 0 \\
1 & 1 \\
\smcq
 & \nu_4 & \nu_4 & \nu_2 & \nu_2 & \nu_6 & \nu_3 & \nu_5 \\
\smaq
1 & 1 \\
0 & 1 \\
\smcq
 & \nu_5 & \nu_5 & \nu_5 & \nu_6 & \nu_3 & \nu_6 & \nu_4 \\
\smaq
1 & 1 \\
1 & 0 \\
\smcq
 & \nu_6 & \nu_6 & \nu_6 & \nu_5 & \nu_4 & \nu_5 & \nu_2 \\
\bottomrule
\end{array}
\ees
\end{center}
\caption{Transformation of the odd characteristics under the action of
 the modular group.}
\label{tab:actodd}
\end{table}

\newpage

\bibliographystyle{unsrt}
\bibliography{bibliografy}
\end{document}